# Quantum and classical correlations in the solid-state NMR free induction decay.

V. E. Zobov[1] and A. A. Lundin[2]

[1]L. V. Kirensky Institute of Physics, Siberian Branch, RAS, 660036, Krasnoyarsk, Russia,
[2]N. N. Semenov Institute of Chemical Physics, RAS, 117977, Moscow, Russia,

e-mail: rsa@iph.krasn.ru

The free-induction decay (FID) of the transverse magnetization in a dipolar-coupled rigid lattice is a fundamental problem in magnetic resonance and in the theory of many-body systems. As it was shown earlier the FID shapes for the systems of classical magnetic moments and for quantum nuclear spins ones coincide if there are many quite equivalent nearest neighbors $V$ in a solid lattice. In this paper, we reduce a multispin density matrix of above system to a two-spin matrix. Then we obtain analytic expressions for the mutual information and the quantum and classical parts of correlations at the arbitrary spin quantum number $S$, in the high-temperature approximation. The time dependence of these functions is expressed via the derivative of the FID shape. To extract classical correlations for $S>1/2$ we provide generalized POVM measurement using the basis of spin coherent states. We show that in every pair of spins the portion of quantum correlations changes from 1/2 to $1/(S+1)$ when S is growing up, and quantum properties disappear completely only if $S\to\infty$ and not in the case when $V\to\infty$.

## I. INTRODUCTION

Nuclear spin systems observed by nuclear magnetic resonance (NMR) really for a long time and yet now perform a suitable laboratory for studying of physics of nonequilibrium processes in quantum many-body systems. Some of the most of fundamental lines of that type activities are the emergence and growth of correlations, spin dynamics and so on [1]. Quite recently applications of the NMR spin dynamics to investigate quantum information processing were initiated [2]. It is usually assumed that the quantum correlation existing both at low and at high temperatures influence the performance speed of quantum computer [3]. In this regard, the interest of researchers has shifted from the calculation of the correlation function as a whole to their partitioning into quantum and classical parts (e.g. see the review [3]). On the other hand different time correlation functions determine observed NMR signals in conventional NMR [1]. However their decomposition in quantum and classical components has not been done yet. In the present article it will be done for one of the most significant NMR time correlation functions, namely, for the free induction decay (FID) function.

The FID shape links to the shape of NMR absorption line via Fourier transform [1]. In the many-body spin systems of solids, the calculation of the time correlation functions is a very challenging problem and different approaches to it solution has been widely discussed. In light of the above, point the works [4, 5]. In the article [4], the numerical simulation has been used to derive FID curves for a simple cubic lattice with 216 classical magnetic moments (classical spins) coupled by dipole-dipole interaction. It was found that the calculated FID shape is close to the FID shape of fluorine nuclei (nuclear spin $S = 1/2$) which was experimentally measured in CaF$_2$ [6]. In Ref. [5] we explained this result. We showed that the time dependence of FID for the system consisting of quantum spins and one formed by classical magnetic moments $\mu = \gamma\hbar\sqrt{S(S+1)}$ coincides in a limit of a large number of the equivalent nearest neighbors surrounding a probe spin (anyone spin) in a lattice. The deduction has been made on the basis of the analysis of various contributions to the spectral moments of all orders of NMR spectrum. Actually in Ref.[5] we demonstrated that if the numbers of the rather equivalent nearest neighbors for any spin is large enough then the principal contributions to the arbitrary NMR spectral moment carrying in by the terms of the moment with maximum number of the summing indexes on the lattice. Referred above contributions coincide exactly for classical and quantum spin systems. So it works for ordinary regular three-dimensional lattices (e.g. simple cubic one). Comparison of the values of the exact spectral moments from $M_4$ to $M_8$, performed in Ref. [5] also revealed insignificant discrepancies between results for the systems of quantum and classical spins.

It is interesting to calculate share of quantum correlations under these conditions. One of the approaches to solution the problem of clearing quantum effects (quantum correlations) consists of the reduction of the multispin density matrix to the two-spin matrix with the subsequent analysis of pair correlations [3]. Thus, such approach is applied to the description of one-dimensional XY-chain in Refs. [7, 8], and also, in Ref. [9], to investigation of spins in nanopore with equal dipolar interaction between any two spins. In both cases, only nuclei with a spin quantum number $S$=1/2 only were studied. In the present work we consider lattices formed by nuclei with an arbitrary spin $S$. Any disturbing quadrupole effects are neglected. We will provide a reduction of the multispin density matrix to a two-spin matrix. Then, following the program put forward in Ref. [10], we are going to calculate shares of quantum and classical correlations: for $S$=1/2 we shall use the von Neumann orthogonal measurement, whereas for $S$>1/2 we shall provide generalized POVM measurement (positive-operator-valued-measure) [3, 11] using the basis of spin coherent states (SCS) [12]. In spite of our basic goal for the present paper consisting of studying spin systems coupled by a dipole-dipole interaction we are going also to consider model lattices with spin-spin interaction only between spin components, parallel to the external magnetic field (Ising like interaction) because the last one allows to get some exact results.

## II. HAMILTONIAN AND SOME BASIC EQUATIONS OF THE PROBLEM.

In traditional experiments employing NMR, the spin temperature considerably exceeds the energy of the Zeeman and other interactions in the spin system. As a consequence, polarization is very small for nuclear spin in the strong static magnetic field at room temperature $T$, $\beta = \hbar\omega_0 / kT \approx 10^{-5} \ll 1$ ($\omega_0$ is the Larmor frequency), and the equilibrium density matrix has the form [1]:

$$\hat{\rho}_{eq} = (1 + \beta \hat{S}_z)/Z, \qquad (1)$$

where Z is the partition function, $\hat{S}_\alpha = \sum_j \hat{S}_{j\alpha}$, $\hat{S}_{j\alpha}$ is the α-component (α = x, y, z) of the spin $j$, and the external magnetic field $H_0$ is directed along the $z$ axis. As well known [1] for observation of a FID signal it is necessary preliminary to prepare the spin system using the pulse of the radio-frequency magnetic field causing rotation of spins at π/2 -angle around the $y$ axis of the rotating with the Larmor frequency reference frame . So we get

$$\hat{\rho}(0) = \hat{Y} \hat{\rho}_{eq} \hat{Y}^{-1} = (1 + \beta \hat{S}_x)/Z.$$

This initial density matrix evolves in time as

$$\hat{\rho}(t) = \hat{U}(t)\hat{\rho}(0)\hat{U}^{-1}(t) = [1 + \beta \hat{U}(t)\hat{S}_x \hat{U}^{-1}(t)]/Z = [1 + \beta \Delta\hat{\rho}(t)]/Z, \qquad (2)$$

where $\hat{U}(t) = \exp(-i\hat{H}t/\hbar)$ is the operator of evolution with the Hamiltonian $\hat{H}$. An observable signal of FID is proportional to time-correlation function:

$$F(t) = \frac{Tr\{\hat{S}_x \hat{\rho}(t)\}}{Tr\{\hat{S}_x \hat{\rho}(0)\}} \qquad (3)$$

and it links to the shape of NMR absorption line via the Fourier transform.

As it is known [1], in nonmetallic diamagnetic solids (at least consisting of light nuclei (e.g., protons or $^{19}$F nuclei)), a principal cause of the absorption NMR line broadening is a secular part of dipole-dipole interaction between nuclear spins. So this interaction completely specifies the dynamics of the nuclear spin system:

$$\hat{H}_d = \sum_{i \neq j} b_{ij} \hat{S}_{zi} \hat{S}_{zj} + \sum_{i \neq j} a_{ij}(\hat{S}_{xi}\hat{S}_{xj} + \hat{S}_{yi}\hat{S}_{yj}) = \sum_{i \neq j} b_{ij}\hat{S}_{zi}\hat{S}_{zj} + \sum_{i \neq j} a_{ij}\hat{S}_{+i}\hat{S}_{-j} = \hat{H}_{zz} + \hat{H}_{ff}, \qquad (4)$$

where $\hat{S}_{i\pm} = \hat{S}_{ix} \pm i\hat{S}_{iy}$, $b_{ij} = \gamma^2 \hbar(1 - 3\cos^2\theta_{ij})/2r_{ij}^3$, $a_{ij} = -b_{ij}/2$, $\vec{r}_{ij}$ is the vector connecting spins $i$ and $j$, $\theta_{ij}$ is the angle, formed by vector $\vec{r}_{ij}$ with the static external magnetic field. From here on, energy is expressed in frequency unities.

Let's suppose that the system is in equilibrium in the strong external magnetic field for which Zeeman splitting substantially exceeds spin-spin interaction (4). Therefore the initial state of the system is well described by the density matrix (1). In this initial state all correlations are absent. In the course of evolution to the state described by Eq. (2) dynamic correlations are forming in the system. One of approaches to their examination consists in a reduction of the multispin density matrix (2) to the two-spin matrix with the subsequent analysis of pair correlations and to their partitioning into quantum and classical parts [3, 7 - 9]. For such a reduction we will choose two spins in the lattice points $i$ and $j$ and then calculate a trace in Eq. (2) over all other spin variables. The density matrix $\hat{\rho}_{ij}(t)$ obtained will depend only on spin states of two nuclei $i$ and $j$ and in the present section we will use appropriate numbers accordingly 1 and 2.

The information-theoretic measure of correlations between two spins is the mutual information [3, 11],

$$I(\hat{\rho}_{12}) = S_N(\hat{\rho}_1) + S_N(\hat{\rho}_2) - S_N(\hat{\rho}_{12}), \tag{5}$$

where $S_N(\hat{\rho}) = -Tr\{\hat{\rho}\log_2 \hat{\rho}\}$ is the von Neumann entropy, $\hat{\rho}_1 = Tr_2\hat{\rho}_{12}$, $\hat{\rho}_2 = Tr_1\hat{\rho}_{12}$ are the density matrices reduced to one spin. We assume to calculate the von Neumann entropy in the lowest order on $\beta$ [1, 10],

$$S_N(\hat{\rho}) = -Tr\{\hat{\rho}\log_2 \hat{\rho}\} \approx \log_2 Z - \frac{\beta^2}{2Z\ln 2}Tr(\Delta\hat{\rho})^2.$$

In the high-temperature approach accepted the mutual information (5) is as follows:

$$I(\hat{\rho}_{12}) = \frac{\beta^2}{2\ln 2}\left\{\frac{1}{d^2}Tr(\Delta\hat{\rho}_{12})^2 - \frac{1}{d}Tr_1(\Delta\hat{\rho}_1)^2 - \frac{1}{d}Tr_2(\Delta\hat{\rho}_2)^2\right\}, \tag{6}$$

where $d = 2S+1$.

The mutual information (5) is used to measure the total correlations, which are sums of the classical and quantum correlations. The classical correlations can be calculated by the measurement, described in [3]. To perform a von Neumann measurement we must project the state $\hat{\rho}_{12}(t)$ on the complete basis of orthogonal wave functions $|\Psi_m\rangle$ by means of a complete set of projectors,

$$\hat{\Pi}_m = |\Psi_m\rangle\langle\Psi_m|, \quad \sum_m \hat{\Pi}_m = 1. \tag{7}$$

In the case of system with $S=1/2$ the complete set of orthogonal projectors of the first spin consists of two projectors of a general form,

$$\hat{\Pi}_{1\pm} = \frac{1}{2}[1\pm(n_x\hat{\sigma}_{1x} + n_y\hat{\sigma}_{1y} + n_z\hat{\sigma}_{1z})], \tag{8}$$

where $n_\alpha$ are the direction cosines, $\hat{\sigma}_\alpha$ are the Pauli matrices, and $\alpha = x, y, z$.

The density matrix $\hat{\rho}_{12}(t)$ is transformed after projecting on the states of the first spin to

$$\hat{\Pi}_1(\hat{\rho}_{12}) = \frac{1}{Z}[1 + \beta\hat{\Pi}_1(\Delta\hat{\rho}_{12}(t))], \tag{9}$$

where we have

$$\hat{\Pi}_1(\Delta\hat{\rho}_{12}(t)) = \sum_m (\hat{\Pi}_{1m} \otimes \hat{E}_2)\Delta\hat{\rho}_{12}(t)(\hat{\Pi}_{1m} \otimes \hat{E}_2),$$

and where $\hat{E}_2$ is the unit matrix.

If one wants to use the generalized POVM measurement, he must recognize that the functions $|\Psi_m\rangle$ in operators (7) can now be no orthogonal, and these operators strictly speaking are then already not projectors [11]. It is assumed that the spin coherent states (SCS) (Bloch states) [12]

$$|\theta,\varphi\rangle = \hat{R}(\theta,\varphi)|S\rangle = \sum_{m=-S}^{m=S} \binom{2S}{S+m}^{1/2} (\cos\theta/2)^{S+m}\left(e^{i\varphi}\sin\theta/2\right)^{S-m}|m\rangle, \tag{10}$$

are closest to the states of the classical momenta. Here $\theta$ and $\varphi$ are the polar and azimuthal angles on the unit sphere (Bloch sphere), $|m\rangle$ is an eigenstate of the operator $S_z$ with eigenvalues $m$ assuming 2S+1 values,

-S, -S+1, … , S-1, S.

These states (10) are obtained from the ground state $|S\rangle$ by the rotation operator $\hat{R}(\theta,\varphi)$ and are a superposition of states with different $m$. The average values of spin projections in the state (10) are as follows

$$\langle\theta,\varphi|\hat{S}_z|\theta,\varphi\rangle = S\cos\theta, \quad \langle\theta,\varphi|\hat{S}_x|\theta,\varphi\rangle = S\sin\theta\cos\varphi, \quad \langle\theta,\varphi|\hat{S}_y|\theta,\varphi\rangle = S\sin\theta\sin\varphi$$

and are the same as for classical momentum. The completeness property

$$\frac{2S+1}{4\pi}\int|\theta,\varphi\rangle\langle\theta,\varphi|\sin\theta d\theta d\varphi = 1$$

is satisfied for the SCS basis, but this basis is not orthogonal.

We take the SCS system as the measurement basis in Eq. (7), to perform the POVM measurement of the first spin, which reduces to multiplying by the SCS and calculating the trace, and obtain the classical density function for the probability distribution of the angle values

$$\hat{\rho}_2(\theta_1\varphi_1;t) = \frac{(2S+1)}{4\pi} Tr_1\{\hat{\rho}_{12}(t)|\theta_1,\varphi_1\rangle\langle\theta_1,\varphi_1| \otimes \hat{E}_2\} = \frac{(2S+1)}{4\pi}\langle\theta_1,\varphi_1|\hat{\rho}_{12}(t)|\theta_1,\varphi_1\rangle. \tag{11}$$

Now to calculate the Shannon entropy we must calculate the integral over the Bloch sphere

$$S_{ShN}(\hat{\rho}_2(\theta_1\varphi_1;t)) = -\int Tr_2\{\hat{\rho}_2(\theta_1\varphi_1;t)\log_2\hat{\rho}_2(\theta_1\varphi_1;t)\}\sin\theta_1 d\theta_1 d\varphi_1.$$

As usual let us choose the mutual information $I(\hat{\Pi}_1(\hat{\rho}_{12}))$ calculated using formulas (5), (9) and (11) for this matrix, as a measure of classical correlations. Unfortunately the gained value will depend on the chosen basis (7). It was proposed [3] to search all bases and to take the maximum value of correlation $I(\hat{\Pi}_1(\hat{\rho}_{12}))$ as the universal measure. However, such a program can be realized only for some simple cases, e.g. for two-level system. If we subtract the classical part from all correlations (5), then we obtain the quantum part of the correlations

$$Q_{12} = I(\hat{\rho}_{12}) - I(\hat{\Pi}_1(\hat{\rho}_{12})). \tag{12}$$

After carrying out of minimization of this quantity on measurement bases one gains an entropy measure of quantum correlations named by quantum discord $D_{12}$ [3]. Measure (12) without optimization was called a measurement dependent discord [3].

### III. THE MODEL CALCULATIONS WITH ISING LIKE-INTERACTION ONLY.

We are going to study the general case of Hamiltonian (4) in the following section, while now, at the first stage, let's put $a_{ij} = 0$. In this case, the time evolution of the matrix (2) can be written out in the explicit form

$$\hat{\rho}(t) = \frac{1}{Z}\left\{1 + \frac{\beta}{2}\left[\sum_i \hat{S}_{i+} \prod_{j(\neq i)} \exp(-it2b_{ij}\hat{S}_{jz}) + \sum_i \hat{S}_{i-} \prod_{j(\neq i)} \exp(it2b_{ij}\hat{S}_{jz})\right]\right\}. \tag{13}$$

So the observable FID signal (3) is

$$F_{zz}(t) = \prod_{j(\neq i)} \frac{\sin(db_{ij}t)}{d\sin(b_{ij}t)}. \tag{14}$$

If a number of equivalent nearest neighbors $V$ in formula (14) is large enough it can be adequate approximated by the Gaussian function

$$F_G(t) = \exp\{-M_2^{zz}t^2/2\} \tag{15}$$

where

$$M_2^{zz} = \frac{4}{3}S(S+1)\sum_j b_{ij}^2.$$

Under these conditions the FID shape does not depend on $S$ and, therefore, coincides with the FID shape of the of system of classical magnetic moments with $\mu = \gamma\hbar\sqrt{S(S+1)}$ which one gets in the limit $S \to \infty$.

We can discriminate functions given by Eqs. (15) and (14) using discrepancy of their fourth moments:

$$M_{G4} = 3(M_2^{zz})^2 = 3\left[\frac{4}{3}S(S+1)\sum_j b_{ij}^2\right]^2,$$

$$M_4^{zz} = 3(M_2^{zz})^2 - \frac{3}{5}\left[\frac{4}{3}S(S+1)\right]^2\left\{2 + \frac{1}{S(S+1)}\right\}\sum_j b_{ij}^4. \tag{16}$$

As it follows from Eq. (16) the above discrepancy $\Delta M_4 = M_{G4} - M_4^{zz}$ expressed through the lattice sum with only one summation via lattice sites whereas quantity of the moment (16) defines by the lattice sums with two such indexes of summation containing in $(M_2^{zz})^2$, and therefore $\Delta M_4 / M_4^{zz} \sim 1/V$.

What does it really means? Whether coincidence of the FID shapes at $V \to \infty$ means disappearance of quantum correlations? To answer this question, we will execute a reduction of a density matrix (13) [7-9]. Let us fix two spins in sites $i$ and $j$ and then calculate a trace in Eq. (13) over all other spin variables. So we get

$$\hat{\rho}_{ij}(t) = \frac{1}{d^2}\left\{1 + \frac{\beta}{2}\left[G_{i(j)}(t)\hat{S}_{i+}\exp(-it2b_{ij}\hat{S}_{jz}) + \right.\right.$$
$$\left.\left. + G_{i(j)}(t)\hat{S}_{i-}\exp(it2b_{ij}\hat{S}_{jz}) + G_{j(i)}(t)\hat{S}_{j+}\exp(-it2b_{ij}\hat{S}_{iz}) + G_{j(i)}(t)\hat{S}_{j-}\exp(it2b_{ij}\hat{S}_{iz})\right]\right\},$$

$$G_{i(j)}(t) = \prod_{f(\neq i,j)} \frac{\sin(db_{if}t)}{d\sin(b_{if}t)}, \quad G_{j(i)}(t) = \prod_{f(\neq i,j)} \frac{\sin(db_{jf}t)}{d\sin(b_{jf}t)}. \tag{17}$$

To simplify the analysis let us suppose that all the spins occupy equivalent positions in the lattice. As a result one can write

$$G_{i(j)}(t) = G_{j(i)}(t) \equiv G_{ij}(t),$$

and the density matrix becomes

$$\hat{\rho}_{ij}(t) = \{1 + \beta \Delta \hat{\rho}_{ij}(t)\}/d^2 \tag{18}$$

where

$$\Delta \hat{\rho}_{ij}(t) = G_{ij}(t)[\hat{S}_{i+}\exp(-it2b_{ij}\hat{S}_{jz}) + \hat{S}_{i-}\exp(it2b_{ij}\hat{S}_{jz}) + $$
$$+ \hat{S}_{j+}\exp(-it2b_{ij}\hat{S}_{iz}) + \hat{S}_{j-}\exp(it2b_{ij}\hat{S}_{iz})]/2.$$

Expression (18) differs from the similar expression for isolated pair of spins, obtained in Ref. [10], owing to replacement of $\tau$ by $t2b_{ij}$ and $\beta$ by $\beta G_{ij}(t)$. Therefore, omitting intermediate evaluations, let's state the final results at once. At first, for the mutual information (6) with $\Delta \hat{\rho}_{12} = \Delta \hat{\rho}_{ij}(t)$ we get

$$I(\hat{\rho}_{ij}) = \frac{(\beta G_{ij}(t))^2}{3\ln 2} S(S+1)[1 - g_{ij}^2(t)], \tag{19}$$

where

$$g_{ij}(t) = \frac{\sin(db_{ij}t)}{d\sin(b_{ij}t)}.$$

And secondly, if $S = 1/2$ we obtain for classical $C_{ij} = I(\hat{\Pi}_1(\hat{\rho}_{ij}))$ and quantum (by using quantum discord (12) $D_{ij}$) parts of correlations

$$C_{ij} = D_{ij} = \frac{1}{2}I(\hat{\rho}_{ij}) = \frac{(\beta G_{ij}(t))^2}{8\ln 2}\sin^2(tb_{ij}). \tag{20}$$

The result (20) is gained by means of the von Neumann orthogonal measurement (9) with the projectors (8) on one of the spins. At last, if $S > 1/2$ for classical ($J_{ij}$) and quantum ($Q_{ij}$) parts of correlations one gets

$$J_{ij} = \frac{(\beta G_{ij}(t))^2}{6\ln 2}\left\{S(S+1)\left[f_{ij}(t) - g_{ij}^2(t)\right] + S^2\left[1 - g_{ij}^2(t)\right]\right\}, \tag{21}$$

$$Q_{ij} = I(\hat{\rho}_{ij}) - J_{ij} = \frac{(\beta G_{ij}(t))^2}{6\ln 2}\left\{S(S+1)\left[1 - f_{ij}(t)\right] + S\left[1 - g_{ij}^2(t)\right]\right\}. \tag{22}$$

In the above equations we use notation

$$f_{ij}(t) = \sum_{n=0}^{n=2S}\binom{2S}{n}\frac{(2n)!!}{(2n+1)!!}(-1)^n(\sin tb_{ij})^{2n}.$$

The formula (21) is gained by using the generalized POVM measurement (11) with the basis from SCS (10).

The expressions (21) and (22) describe evolution of required parts of correlations with time. To make a qualitative analysis of their behavior for the large number of neighbors $V$ we need to pay attention the fact that in this case function $G_{ij}(t)$ (17) rapidly dies out at time scale with the order of $1/\sqrt{M_2^{zz}}$. At such times we have

$$|b_{ij}t| \sim \sqrt{b_{ij}^2/M_2^{zz}} \sim 1/\sqrt{V} \ll 1.$$

Therefore in Eqs. (19), (21) and (22) it is possible to keep only first nonvanishing terms in expansion of the functions $f_{ij}(t)$, $g_{ij}(t)$ and also $\sin tb_{ij}$ in powers of $t$. So we get

$$I(\hat{\rho}_{ij}) \approx \frac{(\beta G_{ij}(t))^2}{9\ln 2}4[S(S+1)b_{ij}t]^2, \tag{23}$$

$$Q_{ij} \approx \frac{(\beta G_{ij}(t))^2}{9\ln 2}4[Sb_{ij}t]^2(S+1). \tag{24}$$

From here for the relative share of quantum correlation we can extract

$$Q_{ij}/I(\hat{\rho}_{ij}) \approx 1/(S+1), \tag{25}$$

i.e. we reveal that as $S$ is growing up the share of quantum correlations decreases. We would note also that if $S = 1/2$ expression (25) is equal to 2/3 whereas from Eq. (20) one gets 1/2. The discrepancy is related to the distinctions in the methods of measurement.

## IV. **Dipole-dipole interaction case**

Now let us study system with the total Hamiltonian (4). Interaction between transversal spin components does not allow writing down just now an explicit time dependence of the density matrix in a so simple form as Eq. (13). In this situation for finding the appropriate form of the density matrix, we shall decompose it over the complete system of orthogonal operators $|k)$ following the line outlined in ref. [13 - 15]. In this representation

$$\hat{S}_x(t) = \hat{U}(t)\hat{S}_x\hat{U}^{-1}(t) = \sum_{k=0}^{\infty} A_k(t)|k). \qquad (26)$$

The initial operator $|0) = \hat{S}_x$. Each subsequent operator of the basis is obtained from the previous one after the procedure of commutation with the Hamiltonian according to the recursion relations:

$$|1) = i[\hat{H}_d, |0)], \quad |k+1) = i[\hat{H}_d, |k)] + v_{k-1}^2|k-1) \quad (if \ k \geq 1), \qquad (27)$$

$$v_k^2 = Sp\{(k+1|k+1)\}/Sp\{(k|k)\}.$$

For amplitudes $A_k(t)$ the system of the differential equations [13, 14] has been revealed

$$\dot{A}_0(t) = v_0^2 A_1(t), \quad \dot{A}_k(t) = A_{k-1}(t) - v_k^2 A_{k+1}(t) \quad (if \ k \geq 1). \qquad (28)$$

To avoid confusion, a certain difference in the definition of amplitudes $A_k(t)$ in references [13] and [14] should be noticed. The difference is in the factor $(i)^k$. We have chosen a variant used in ref. [14] at which functions $A_k(t)$ contain no imaginary part, because the factor $(i)^k$ is included into definition of operators $|k)$. The parameters $\{v_k\}$ which values determine the solution of the system (28), are expressed unequivocally through the moments of the NMR absorption line [13]. In particular

$$v_0^2 = M_2 = 3S(S+1)\sum_j b_{ij}^2, \quad v_1^2 = (M_4 - M_2^2)/M_2, \quad v_2^2 = (M_2 M_6 - M_4^2)/(M_4 - M_2^2)M_2, \qquad (29)$$

where $M_2, M_4, M_6$ are the second, fourth and sixth moments of the NMR absorption line.

Let us substitute decomposition (26) to Eq. (2) and then execute the reduction. As it means we have to choose two spins at sites $i$ and $j$ and then to calculate a trace in Eq. (2) over all other spin variables. Thus we have

$$\hat{\rho}_{ij}(t) = \frac{1}{d^2}\left\{1 + \beta\sum_{k=0}^{\infty} A_k(t)\frac{d^2}{Z}\underset{\neq i,j}{Tr}[k)\right\}. \qquad (30)$$

So for the first two orthogonal operators of the complete set we get

$$\frac{1}{Z}\underset{\neq i,j}{Tr}|0) = \frac{1}{Z}\underset{\neq i,j}{Tr}\sum_f \hat{S}_{xf} = \frac{1}{d^2}(\hat{S}_{xi} + \hat{S}_{xj}), \qquad (31)$$

$$\frac{1}{Z}\underset{\neq i,j}{Tr}[1] = \frac{i}{Z}\underset{\neq i,j}{Tr}[\hat{H}_d,\hat{S}_x] = \frac{-2}{d^2}(b_{ij}-a_{ij})(\hat{S}_{yi}\hat{S}_{zj}+\hat{S}_{yj}\hat{S}_{zi}). \tag{32}$$

The contribution to Eq. (30) from orthogonal operators of the higher order can be obtained in two cases. First case assumes zero direct interaction between the chosen spins $i$ and $j$. It is a possible case for example, if the angle $\theta_{ij}$ between the vector $\vec{r}_{ij}$ and external magnetic field is equal to the "magic" value $54^0 44$`. In this situation we have to take into account the contribution from vector $|3\rangle$ which depends on the constant $b_{if}b_{jf}^2$ through the intermediate spin $f$ if this constant is distinct from zero. The second case appears if $S > 1/2$ because orthogonal operators of the high order are formed of products of spin operators not only from different sites, but also from the same site. For examples in vector $|2\rangle$ there is a contribution $\hat{S}_{xi}\{\hat{S}_{zj}^2 - S(S+1)/3\}$, and in vector $|3\rangle$ one gets a contribution $\hat{S}_{yi}\{\hat{S}_{zj}^3 - \hat{S}_{zj}(3S^2+3S-1)/5\}$. We shall neglect above mentioned contributions in Eq. (30) as far as these parts do not contain new qualitative properties, and are small corrections to contributions from Eqs. (31) and (32). The trifle of discussing corrections is a consequence of the different time dependence of the different order amplitudes: $A_k(t) \sim t^k$ at small times. Because of the rapid decay of amplitudes at times $t \geq 1/\sqrt{M_2}$, each additional power of $t$ adds only a small factor $|b_{ij}t| \sim \sqrt{b_{ij}^2/M_2} \sim 1/\sqrt{V} \ll 1$.

Having retained two contributions (31) and (32) in Eq. (30) we get

$$\hat{\rho}_{ij}(t) \approx \frac{1}{d^2}\{1 + \beta A_0(t)(\hat{S}_{xi}+\hat{S}_{xj}) + \beta A_1(t)B_{ij}(\hat{S}_{yi}\hat{S}_{zj}+\hat{S}_{yj}\hat{S}_{zi})\}, \tag{33}$$

where $B_{ij} = -2(b_{ij}-a_{ij}) = -3b_{ij}$ for the Hamiltonian (4). At last, at the further reduction to one spin one gets

$$\hat{\rho}_{i(j)}(t) \approx \frac{1}{d}\{1 + \beta A_0(t)\hat{S}_{xi(j)}\}. \tag{34}$$

Having substituted Eq. (34) in Eq. (3), we get $F(t) = A_0(t)$.

The density matrix (33) looks like similar expression for isolated pair of the spins, calculated in [10] at small times. Therefore, skipping on intermediate evaluations, we are giving the results at once. By such a way we calculated for the mutual information

$$I(\hat{\rho}_{ij}) \approx \frac{\beta^2}{9\ln 2}[S(S+1)B_{ij}A_1(t)]^2 = \frac{\beta^2 b_{ij}^2}{M_2^2 \ln 2}[S(S+1)\dot{F}(t)]^2. \tag{35}$$

Under the transformations in process of obtaining the Eq. (35) formulas (28) and (29) were used. We obtain also that the quantum discord $D_{ij}$ (if S=1/2) and the quantum part of correlations $Q_{ij}$ (if $S > 1/2$) are related to the mutual information $I(\hat{\rho}_{ij})$ from Eq. (35) by the same relations (20) and (25), as in the previous example:

$$C_{ij} = D_{ij} = I(\hat{\rho}_{ij})/2, \qquad Q_{ij} \approx I(\hat{\rho}_{ij})/(S+1).$$

On the basis of the results derived above it can be concluded that the time dependence of the mutual information (35) and the quantum part of correlations is revealing through the derivative of FID shape. Thus rapid exhaustion of pair correlations and reduction of their peak values with the growing up of the number of neighbors $V$ generally speaking do not mean impairment of correlated relations of spins, but mean redistribution of pair correlations to more complicated multispin ones. As a measure of total correlation the total information [3, 16] can serve:

$$T(\hat{\rho}) = \sum_i S_N(\hat{\rho}_i) - S_N(\hat{\rho}) \approx \frac{\beta^2}{3\ln 2} S(S+1)[1 - A_0^2(t)]. \tag{36}$$

At the initial moment of time $A_0^2(0) = 1$ and $T(\hat{\rho}) = 0$. For a long times $A_0^2(t)$ is coming to zero and therefore $T(\hat{\rho})$ reaches own limiting value only defined by entry conditions: e.g. by polarization $\beta$ at given temperature and at the fixed strength of the external magnetic field.

## V. Conclusion

Our results mean that in spite of coincidence [5] of the FID shapes both of classical and quantum spin systems for a large number $V$ of nearest neighbors, the quantum properties of the system are not lost. For every pair of spins the portion of quantum correlations changes from 1/2 to 1/(S+1) with S growing up. In reality the quantum properties disappear completely only if $S\to\infty$ but not in the case when $V\to\infty$. The similarity of the FID shapes means that measurable classical correlations and "immeasurable" (lost at measurement) quantum correlations are bringing the equal influence at FID. So it implies that unobservable simultaneously spin components $\hat{S}_x, \hat{S}_y, \hat{S}_z$ are capable to give the contribution to dynamics of spins simultaneously. Thereof the time scale dependence is determined by the quantity $S(S+1)$, instead of $S^2$, where $S$ is the maximal value of an observable projection upon any axis.